\documentclass{ifacconf}

\usepackage{graphicx}      
\usepackage{natbib}        
\usepackage{siunitx}
\usepackage{amsmath}
\usepackage{comment}
\usepackage[%
    font={small,sf},
    labelfont=bf,
    format=hang,    
    format=plain,
    margin=0pt,
    width=0.8\textwidth,
]{caption}
\usepackage[list=true]{subcaption}

\begin{document}
\sloppy
\begin{frontmatter}

\title{Global Sensitivity Methods for Design of Experiments in Lithium-ion Battery Context \thanksref{footnoteinfo}} 

\thanks[footnoteinfo]{X. Xie acknowledges support from the International Max Planck Research School for Advanced Methods in Process and Systems Engineering, MPI~Magdeburg. D.M. Raimondo has been (partially) supported by the Italian Ministry for Research in the framework of the 2017 Program for Research Projects of National Interest (PRIN), Grant no. 2017YKXYXJ.}

\author[Second]{A. Pozzi} 
\author[First]{X. Xie}
\author[Second]{D.M. Raimondo}
\author[First]{R. Schenkendorf}

\address[First]{Institute of Energy and Process Systems Engineering, 
   TU Braunschweig, 38106 Braunschweig, Germany (e-mail: \{x.xie,r.schenkendorf\}@tu-braunschweig.de)}
\address[Second]{Department of Electrical, Computer and Biomedical Engineering,
   University of Pavia, 27100 Pavia, Italy (e-mail: andrea.pozzi03@universitadipavia.it, davide.raimondo@unipv.it)}

\begin{abstract}                
Battery management systems may rely on mathematical models to provide higher performance than standard charging protocols. Electrochemical models allow us to capture the phenomena occurring inside a lithium-ion cell and therefore, could be the best model choice. However, to be of practical value, they require reliable model parameters. Uncertainty quantification and optimal experimental design concepts are essential tools for identifying systems and estimating parameters precisely. Approximation errors in uncertainty quantification result in sub-optimal experimental designs and consequently, less-informative data, and higher parameter unreliability. In this work, we propose a highly efficient design of experiment method based on global parameter sensitivities. This novel concept is applied to the single-particle model with electrolyte and thermal dynamics (SPMeT), a well-known electrochemical model for lithium-ion cells. The proposed method avoids the simplifying assumption of output-parameter linearization (i.e., local parameter sensitivities) used in conventional Fisher information matrix-based experimental design strategies. Thus, the optimized current input profile results in experimental data of higher information content and in turn, in more precise parameter estimates.   
\end{abstract}


\end{frontmatter}

\section{Introduction}

In the field of lithium-ion batteries, first-principle models have proven to be beneficial in providing battery management systems with high performance and high safety standards \citep{Chaturvedi2010}.  First-principle models are used in general to gain physical insights, monitor, and control complex processes. 
  For models to be reliable, high accuracy, in terms of structure and parameters, is required. In practice, model structures are approximated
  based on simplifying assumptions that aim to guarantee identifiability while retaining a physical interpretation.
  Unfortunately, the possible imprecision of the structure, together with the fact that model parameters are obtained in general using noisy measurement data, may result in uncertain parameter estimates and inaccurate simulation results \citep{Walter1997}.
 To alleviate these issues, the use of accurate uncertainty quantification in combination with a model-based design of experiment (MBDoE) can provide an improved model calibration with more reliable parameter estimates.
 In particular, the MBDoE consists of finding the optimal input sequence able to minimize the uncertainty of the model parameters \citep{pukelsheim2006optimal}.
 \par
 Most implementations of MBDoE are based
 on the Fisher information matrix  (FIM), which gives a local measurement of how informative a measured signal is in terms of parameter sensitivity.
 In particular, the FIM represents a perfect MBDoE measure only for linear parameter identification problems  \citep{Kiefer1959,Walter1997,Hahn2017}. Local parameter sensitivities assume a linear relationship between model parameter variations and simulation results.  In the nonlinear case, local sensitivities may lead to biased local parameter sensitivity and MBDoE measures because of biased reference parameters \citep{Gunawan2017}.
Moreover, the actual parameter values for calculating these local parameter sensitivities are unknown, and, in turn, the best parameter estimates have to be used. 
 \par
  In the last decade, the use of global sensitivities for the MBDoE has been discussed in the literature \citep{Fernandez2007,Chu2013,Schenkendorf2018}. 
Global parameter sensitivities, by definition, represent nonlinear and multivariate parameter dependencies adequately. Global parameter sensitivities consider model parameters and simulation results as random variables. Thus, the global sensitivity analysis (GSA) aims to quantify the amount of variation that each model parameter contributes to the variation in the simulation results.
 \par
 Most lithium-ion cell models employed in advanced battery management systems (BMSs) can be grouped into two main categories: equivalent circuit models (ECMs; \cite{hu2012comparative}), which are intuitive and straightforward, and electrochemical models (EMs; \cite{Gomadam2002,Santhanagopalan2006a}), which are far more accurate. The pseudo-two-dimensional (P2D) model \citep{Doyle1993}, which consists of nonlinear partial differential algebraic equations (PDAEs),  is the most widely used EM. However, the use of the latter for control purposes is  limited due to its high computational burden and  its identifiability and observability issues \citep{forman2012genetic,Moura2015}.  
The aforementioned issues can be addressed by reduced order models, which have  raised the interest of the research community, due to the fact that they still provide a sufficiently detailed description of the electrochemical phenomena \citep{Zou2014}. Parameter identifiability and state observability of the single particle model (SPM;  \cite{Ning2004,Santhanagopalan2006a}), which models the electrodes as single particles, have been analyzed in several works \citep{DiDomenico2010,bizeray2018identifiability,Pozzi2018c}. 
The electrolyte  (SPMe; \cite{Moura2017}) and thermal dynamics (SPMeT;  \cite{Perez2016}) can also be considered in order to increase the model accuracy.
 In the literature, various  MBDoE studies exist, including ones for complex (electro)chemical processes. For instance, the usefulness of MBDoE for lithium-ion battery models (both full and reduced-order ones) was demonstrated recently in \cite{mendoza2016optimization, mendoza2017maximizing, Pozzi2018, park2018a, park2018b}, where the current profile is optimized by relying on the FIM to maximize the identifiability of the parameters.
\par
 To the best of the authors' knowledge, the usefulness of GSA and its effective implementation for (electro)chemical processes in the context of MBDoE have not been analyzed thus far.
In this work, we propose a highly efficient MBDoE framework that is based on GSA, and we implement this novel GSA-MBDoE concept for a lithium-ion cell modelled as SPMeT. It has to be noticed that the proposed concept avoids the simplifying assumption of output-parameter linearization used in standard MBDoE strategies. To prevent a computation overload while replacing the local parameter sensitivity matrix and solving the underlying dynamic optimization problem, we make use of the point estimate method (PEM) as a highly efficient sampling technique to determine global parameter sensitivities. In contrast to previous work \citep{Schenkendorf2018}, the global parameter sensitivities are directly transferred to standard DoE criteria.
\par
The paper is organized as follows. In Section 2, the  SPMeT is introduced, including modeling assumptions and governing equations. In Section 3, the basic concepts of the model-based design of experiments are summarized, and the novel experimental design, which is based on global parameter sensitivities, is proposed. In Section 4, the standard MBDoE approach and the novel GSA-MBDoE concepts are applied to the lithium-ion battery model and critically compared. Finally, in Section
5, the results are summarized and conclusions provided.    
\section{Lithium-ion Battery Model}

In this study, we consider the  single-particle model with electrolyte and thermal dynamics, which has  proven to be  accurate enough but also suitable for real-time implementation in advanced battery management systems \citep{Moura2017}.  
In the following, the cell sections are indexed with  $j \in \{p,s,n\}$ in all the equations except for those valid only for the electrodes where the index $i$ refers to $\{p,n\}$  instead. The variables $t\in \mathbf{R}$, $x \in \mathbf{R}$ and $r \in \mathbf{R}$ indicate respectively the time index, the spatial direction along which the lithium ions are transported and the radial distance within an active particle at location x.  As previously done in this context by \cite{Subramanian2005}, a fourth-order polynomial approximation  of the ion concentration along the radial axis $r$ of each electrode is considered. In particular, the concentration is described as a function of  $r$, whose coefficients depend on the solid average concentration $\bar c_{s,i}(t)$ and the average concentration flux $\bar q_{i}(t)$. Let  the average stoichiometry in the electrodes be defined by:
\begin{align}
\bar \theta_i (t)=\frac{\bar c_{s,i}(t)}{c_{s,i}^{max}},
\end{align}
where $c_{s,i}^{max}$ is the maximum solid concentration. 
The dynamics of the average stoichiometries can be expressed by the following equation \citep{Subramanian2005}:
\begin{align}
\dot{\bar \theta}_i(t)=\frac{3 I_{app}(t)}{a_i R_{p,i} L_i F A  c_{s,p}^{max}},
\end{align}
where the thickness of the $i$th section is described by $L_i$, the particle radius is denoted by $R_{p,i}$, $F$ is the Faraday constant, $A$ is the  area of the cell, $I_{app}(t)$ is the input current  (with the assumption that a negative current charges the cell) and $a_i=\frac{3 \epsilon_i^{act}}{R_{p,i}}$ is the specific active surface area, with $\epsilon_i^{act}$ the active material volume fraction  defined by:
\begin{subequations}
\begin{align}
\epsilon^{act}_p&=-\frac{C}{\Delta \theta_p A F L_p c_{s,p}^{max}},\\
\epsilon^{act}_n&=\frac{C}{\Delta \theta_n A F L_n c_{s,n}^{max}},
\end{align}
\end{subequations}
in which the cell capacity is represented by $C$. Considering $\theta_{i}^{0\%}$ and $\theta_{i}^{100\%}$ the stoichiometries of the electrodes in the case of fully discharged and fully charged cell respectively, it holds that  $\Delta \theta_i= \theta_i^{100\%}- \theta_i^{0\%}$.
The concentration fluxes present the following dynamics: 
\begin{subequations}\label{eq:average_concentration_flux}
\begin{align}
	\dot {\bar q}_p(t) &= -30 \frac{D_{s,p}(T(t))}{R_{p,p}^2}\bar q_p(t) + \frac{45}{2R_{p,p}^2 F A L_p a_p}I_{app}(t),\\
		\dot {\bar q}_n(t) &= -30 \frac{D_{s,n}(T(t))}{R_{p,n}^2}\bar q_n(t) - \frac{45}{2R_{p,n}^2 F A L_n a_n}I_{app}(t),
\end{align} 
\end{subequations}
where $ T(t)$ is the temperature, and $D_{s,i}(T(t))$ is the solid diffusion coefficient for the $i$th section, which depends on the temperature according to the Arrhenius law. Considering the assumption of  lithium moles conservation in the solid phase \citep{DiDomenico2010}, the average stoichiometry in the anode can be obtained directly from the cathode as follows:
\begin{align}
\bar \theta_n(t)= \theta^{0\%}_n+\frac{\bar \theta_p(t)-\theta^{0\%}_p}{\theta_p^{100\%}-\theta^{0\%}_p} (\theta_n^{100\%}-\theta^{0\%}_n ).
\end{align}
In this way, $\bar \theta_n (t)$  can be considered  as an output of the system, thus reducing the number of state equations and increasing the computational efficiency. 

The surface stoichiometries in the electrodes can be computed with the following algebraic equations:
\begin{subequations}\label{eq:stoichiometric_surface}
\begin{align}
\theta_p(t)&= \bar \theta_p(t)+\frac{8R_{p,p}\bar q_p(t)}{35c_{s,p}^{max}}+\frac{R_{p,p} I_{app}(t)}{35D_{s,p}(T(t))F A L_p a_p c_{s,p}^{max}},\\
\theta_n(t)&= \bar \theta_n(t)+\frac{8R_{p,n}\bar q_n(t)}{35c_{s,n}^{max}}-\frac{R_{p,n} I_{app}(t)}{35D_{s,n}(T(t))F A L_n a_n c_{s,n}^{max}},  
\end{align}
\end{subequations}
according to the polynomial approximation of the lithium concentration along the radius of the particle \citep{Subramanian2005}.

The  state of charge (SOC) is defined as:
\begin{align}\label{eq:soc}
SOC(t)=100\frac{\bar \theta_{n}(t)-\theta_n^{0\%}}{\theta_n^{100\%}-\theta_n^{0\%}}.
\end{align}
\par
A fundamental output of the SPMeT is the terminal voltage $V(t)$. This latter not only depends on the lithium concentration in the solid phase  but also on the one in the electrolyte. Therefore, the PDAEs governing the diffusion of the electrolyte concentration $c_{e,j}(x,t)$ must be considered. In this work, the finite volume method is exploited for spatially discretizing such PDAEs, as previously done in this context by \cite{Torchio2016}. The authors in \cite{Torchio2016} divide the spatial domain  into $P$ non-overlapping volumes for each section. For each section $j$, each volume ranges within $\Omega_{j,k}=\left[x_{j,\bar{k}},x_{j,\underline{k}}\right]$, with $k=1, \cdots, P$, with center $x_{j,k}$ and width $\Delta x_j=L_j/P$. Defining $c_{e,j}^{[k]}(t)$ as the average electrolyte concentration over the $k$th volume of the $j$th section gives: 
\begin{subequations}
\label{eq:ce}
\begin{align}
\epsilon_p \frac{\partial  c^{[k]}_{e,p}(t)}{\partial t}=&  \left.\left[\frac{\tilde{D}_e(x,T(t))}{\Delta x_p}   \frac{\partial c_{e,p}(x,t)}{\partial x} \right]\right|_{x_{p,\underline{k}}}^{x_{p,\bar{k}}}\hspace{-0.4cm}-\frac{1-t_+}{FAL_p}I_{app}(t),\\
\epsilon_s \frac{\partial  c^{[k]}_{e,s}(t)}{\partial t}=&  \left.\left[\frac{\tilde{D}_e(x,T(t))}{\Delta x_s}   \frac{\partial c_{e,s}(x,t)}{\partial x} \right]\right|_{x_{s,\underline{k}}}^{x_{s,\bar{k}}},\\
\epsilon_n \frac{\partial  c^{[k]}_{e,n}(t)}{\partial t}=& \left.\left[\frac{\tilde{D}_e(x,T(t))}{\Delta x_n}   \frac{\partial c_{e,n}(x,t)}{\partial x} \right]\right|_{x_{n,\underline{k}}}^{x_{n,\bar{k}}} \hspace{-0.4cm}+\frac{1-t_+}{FAL_n}I_{app}(t),
\end{align}
\end{subequations}
where $t_+$ is the transference number, $\epsilon_j$ is the material porosity, and $\tilde{D}_e(x,T(t))$ is the electrolyte diffusion coefficient which is computed according to the harmonic mean. See  \cite{Torchio2016} for a further description of the terms in the electrolyte dynamics. Note that the effective diffusion and conductivity coefficients are according to the Bruggeman's theory, where $\tau_j$ represents the tortuosity factor for each section. 

The dependence on the temperature of the parameters above is described by the Arrhenius law, which, for a generic parameter $\psi(T(t))$, is given by:
\begin{align}
    \psi(T(t))=\psi^0 e^{-\frac{E_{a,\psi}}{RT(t)}},
\end{align}
where $\psi^0$ and  $E_{a,\psi}$ are the  constant coefficient and the activation energy related to the parameter $\psi(T(t))$, and $R$ is the universal gas constant.

The terminal voltage  is then given by:
\begin{align}\label{eq:voltage}
\begin{split}
	V(t) =&  -I_{app}(t)R_{sei}+\bar U_p(t) - \bar U_n(t) + \bar \eta_p(t) - \bar \eta_n(t)\\& +  \Delta\Phi_e(t), 
\end{split}
\end{align}
where $R_{sei}$ is the solid electrolyte interface (SEI) resistance, while $\bar U_p(t)$ and $\bar U_n(t)$ are the Open Circuit Potentials (OCPs) in the positive and negative electrodes. The overpotentials $\bar \eta_p(t)$ and $\bar \eta_n(t)$, for the positive and negative electrodes are given respectively by:
\begin{subequations}\label{eq:overpotentials}
\begin{align}
	\bar \eta_p(t) &= \frac{2RT(t)}{F}\sinh^{-1} \left(\frac{-I_{app}(t)}{2 A L_p a_p \bar i_{0,p}(t)}			\right),  \\
		\bar \eta_n(t) &= \frac{2RT(t)}{F}\sinh^{-1} \left(\frac{I_{app}(t)}{2 A L_n a_n \bar i_{0,n}(t)}			\right).
\end{align}
\end{subequations}
The exchange current density is defined as:
\begin{align}
	\bar i_{0,i}(t) = Fk_i(T(t))\sqrt{\bar c_{e,i}(t)\theta_i(t)(1-\theta_i(t))},
\end{align}
where $k_i(T(t))$ is the temperature-dependent rate reaction coefficient and $\bar c_{e,i}(t)$ is obtained by averaging  the  electrolyte concentration over the $i$th section concentration as follows:
\begin{align}
\bar c_{e,i}(t)=\frac{1}{P}\sum_{k=1}^P c_{e,i}^{[k]}(t).
\end{align} 
Moreover, $\Delta\Phi_e(t)$ is computed as: 
\begin{align}
	\Delta\Phi_e(t) = \Phi_e^{drop}(t) + \frac{2RT}{F}(1-t_+)\log_e{\left(\frac{c_{e,p}^{[1]}}{c_{e,n}^{[P]}}	\right)},
\end{align}
where the shape of the ionic  current $i_e(x,t)$ is assumed to be trapezoidal  over the spatial domain \citep{Moura2017}. The electrolyte voltage drop $\Phi_e^{drop}(t)$ can be approximated by:
\begin{align}
\Phi_e^{drop}(t)\simeq  -\frac{I_{app}(t)}{2 A} \left( \phi_p(t) + 2\phi_s(t) + \phi_n(t) \right),
\end{align}
in which:
\begin{subequations}
\label{eq:phi}
\begin{align}
\phi_p(t)&=\Delta x_p \sum_{k=1}^P \frac{2k-1}{\kappa(c_{e,p}^{[k]}(t))\epsilon_p^{p_p}}, \\
\phi_s(t)&=\Delta x_s \sum_{k=1}^{P} \frac{1}{\kappa(c_{e,s}^{[k]}(t))\epsilon_s^{p_s}},\\
\phi_n(t)&=\Delta x_n \sum_{k=1}^{P} \frac{2P-2k+1}{\kappa(c_{e,n}^{[k]}(t))\epsilon_n^{p_n}},
\end{align}
\end{subequations}
where, for the $k$th volume of the $j$th section,  the  electrolyte conductivity is described by $\kappa(c_{e,j}^{[k]}(t))$. This latter that can be derived empirically is expressed with a nonlinear function of the electrolyte concentration:
\begin{align}\label{eq:conductivity_fun}
\begin{split}
\kappa(\gamma_j^{[k]}(t))=&\Big(0.2667 \left(\gamma_j^{[k]}(t)\right)^3 -1.2983  \left(\gamma_j^{[k]}(t)\right)^2\\& +1.7919 \gamma_j^{[k]}(t) + 0.1726\Big) e^{\frac{-E_{a,\kappa}}{RT(t)}},
\end{split}
\end{align}
where $\gamma_j^{[k]}(t)=10^{-3}c_{e,j}^{[k]}(t)$. The function in Eq. \eqref{eq:conductivity_fun} is taken from  \cite{Ecker2015} as well as the expressions of the OCPs in terms of the surface stoichiometries: 
\begin{subequations}
\begin{align}\label{eq:OCPs_fun}
\begin{split}
\bar U_p(t)=&18.45\theta_p^6(t)-40.7 \theta_p^5(t)+20.94\theta_p^4(t)\\&+8.07\theta_p^3(t)-7.837\theta_p^2(t) + 0.02414\theta_p^1(t)+4.571,
\end{split}
\\
\bar U_n(t)=&    \frac{0.1261\theta_n(t)+0.00694}{\theta_n^2(t)+0.6995\theta_n(t)+0.00405},
\end{align}
\end{subequations}
which are fitted from experimentally collected  data. Note that the empirical functions in Eqs. \eqref{eq:conductivity_fun} and \eqref{eq:OCPs_fun} may vary according to the considered cell (in the presented paper, Kokam SLPB 75106100). 

Finally, the temperature dynamics is given by a lumped thermal model  \citep{Perez2016, Perez2017}:
\begin{align}
C_{th} \dot T(t)= Q(t) - h_{c}A_{c}(T(t)-T_{sink}),
\end{align}
where $C_{th}$ is the  thermal capacity of the  cell, and $h_{c}$ and $A_{c}$ are the convective coefficient  and the area of the heat exchange  with the coolant, respectively. We assume that the coolant temperature is constant and equal to $T_{sink}$.  The heat  $Q(t)$ is generated by the cell polarization as follows:
\begin{align}
Q(t)=\vert I_{app}(t) \vert \cdot \vert V(t) - (\bar U_{p}(t)-\bar U_{n}(t)) \vert . 
\end{align}

\par
The electrochemical parameters adopted  are those measured in \cite{Ecker2015,Ecker2015a}, in which a  commercial cell (the Kokam SLPB 75106100) is completely characterized through experiments. The   value of the sink temperature is constant and set to  $T_{sink}=$ \SI{298.15}{K}, while the thermal capacity is set to $C_{th}=$ \SI{4186}{JK^{-1}}. Finally, the heat exchange parameters are assumed to be $A_{c}=$ \SI{1}{m^2} and $h_{c}=$\SI{10}{\watt\per\square\metre\per\kelvin}.

\section{Model-based Design of Experiments}

Next, we propose the GSA-MBDoE concept. The underlying dynamic optimization problem is introduced first. Then, the basics of local and global parameter sensitivities are briefly summarized. The point estimate method is presented to ensure fast GSA-MBDoE results.

\subsection{Optimization Framework}

In this study, the MBDoE states a dynamic optimization problem and reads as: 
\begin{subequations}\label{eq:opt_str}
\begin{align}
    &{\ }\max\limits_{\mathbf{u}(\cdot)}{\ } \Phi(S^{l/g}(\mathbf{p})) \label{equ:obj}\\
    & \nonumber  \text{subject to:} \\
	&{\mathbf{\dot{x}_d}}(t) = \mathbf{f}(\mathbf{x_d}(t),\mathbf{u}(t),\mathbf{p}), \label{equ:dae_1_1} \\
	&\mathbf{x_d}(t_0) = \mathbf{x}_0, \\
	&0 \leq \mathbf{h_{nq}}(\mathbf{x_d}(t),\mathbf{u}(t),\mathbf{p})\label{equ:ineq}, \\
	&\mathbf{u}_{min} \leq \mathbf{u} \leq \mathbf{u}_{max},
	\end{align}
\end{subequations}
where $t \in [t_0,t_0+t_{exp}]$ is the time, with $t_0=\si 0s$ is the initial time and $t_{exp}$ is the time duration  of the experiment, $\mathbf{u} \in \mathbf{R}^{n_u}$ is the vector of the control variables, $\mathbf{p} \in \mathbf{R}^{n_p}$ is the vector of the time-invariant parameters, and $\mathbf{x_d} \in \mathbf{R}^{n_{x_d}}$
 are the differential states. The~initial conditions for the differential states are given by $\mathbf{x}_0$ while $\mathbf{y} \in \mathbf{R}^{n_y}$ is the vector of the model output. Eq. \eqref{equ:dae_1_1} is the model  equation with $\mathbf{f}: \mathbf{R}^{n_{x_d}\times n_{u} \times n_{p}} \rightarrow \mathbf{R}^{n_{x_d}}$. To satisfy critical process constraints, Eq. \eqref{equ:ineq} represents the inequality constraints $\mathbf{h_{nq}}: \mathbf{R}^{n_{x_d}\times n_{u} \times n_{p}} \rightarrow \mathbf{R}^{n_{nq}}$.  [$\mathbf{u}_{min}$,$\mathbf{u}_{max}$] are the upper and lower boundaries for the  control variables. The parameter sensitivity measure $S^{l/g}(\mathbf{p})$ determines the effectiveness of the MBDoE strategy. In this work, local, $S^{l}$, and global parameter sensitivities, $S^{g}$, are used. The parameter sensitivities are translated to a cost function $\Phi(\cdot)$, where different MBDoE cost functions exist in the literature \citep{Walter1997, pukelsheim2006optimal}.   
 
Let us consider a sampling time $t_s$ and a positive integer number $K=\frac{t_{exp}}{t_s}$ of discrete time measurements $\mathbf{y}^{\text{data}}(t_k)$, with  $t_{k+1}=t_k+t_s$, for $k\in [0,\,K]$. Assuming a maximum likelihood estimation procedure, the actual parameter identification problem reads as:
\begin{equation}
\hat{\mathbf{p}} = \arg \min_{\mathbf{p}} \sum\limits_{k=1}^K || \mathbf{y}^{\text{data}}(t_k) - \mathbf{y}(t_k,\mathbf{p})  ||^2_2, \label{eq:parameterIdentification}
\end{equation}
where $||\cdot||_2$ denotes the Euclidean norm, and the model output equation is defined as:
\begin{equation}
    	\mathbf{y}(t_k,\mathbf{p}) = \mathbf{g}(\mathbf{x_d}(t_k,\mathbf{p}))\label{equ:output},
\end{equation}
with $\mathbf{g}: \mathbf{R}^{{x_d}} \rightarrow \mathbf{R}^{n_{y}}$.
Due to additive measurement noise and the Doob--Dynkin lemma \citep{Rao2006}, the identified model parameters $\hat{\mathbf{p}}$ can be considered to be random variables, where the probability space ($\Omega, \mathcal{F},P$) is defined with the sample space $\Omega$, the~$\sigma$-algebra $\mathcal{F}$, and the probability measure $P$. Precise parameter estimates necessitate, in addition to the high data quality (e.g., low measurement noise), high parameter sensitivities. Consequently, the MBDoE and the GSA-MBDoE aim to maximize parameter sensitivities.

\subsection{Parameter Sensitivities}
\label{sec:paraSens}
In the literature, local sensitivities are the standard in the MBDoE \citep{Turanyi1990,ScireJr.2001,Saltelli2005}. 
Local sensitivities $S^{l}(\mathbf{p})$ are given as:
\begin{equation}
  S^{l}[j,i](t) = \left. \frac{\partial \mathbf{y}[j](t)}{\partial \mathbf{p}[i]}\right|_{\mathbf{\hat{p}}}, \label{eq:localSens}
\end{equation}
where $S^{l}(\mathbf{p}) \in \mathbf{R}^{n_y \times n_p}$, and $\mathbf{\hat{p}}$ is the latest update of the estimated model parameter vector. Note that local parameter sensitivities $S^{l}(\mathbf{p})$ are an essential component of the FIM and the MBDoE \citep{, Walter1997, pukelsheim2006optimal}.   

Alternatively, GSA treats the model parameters, $\mathbf{p}$, and the model outcomes, $\mathbf{y}$, as random variables and aims to quantify the amount of variance that each parameter, ${\mathbf{p}[i]}$, contributes to the total variance of the \textit{j}th model output, ${\sigma^2(\mathbf{y}[j](t))}$ \citep{Saltelli2005}. The conditional variance is given as ${\underset{-i}{\sigma}^2(\mathbf{y}[j](t)|\mathbf{p}[i])}$, and the subscript ${-i}$ indicates that the variance is taken over all parameters other than $\mathbf{p}[i]$. The expected value of the resulting conditional variance reads as $\underset{i}{E}\left[\underset{-i}{\sigma}^2(\mathbf{y}[j](t)|\mathbf{p}[i])\right]$, and the subscript notation of $\underset{i}{E}$ indicates that the expected value is taken only over the parameter ${\mathbf{p}[i]}$. The total output variance, ${\sigma^2(\mathbf{y}[j](t))}$, is split into two additive terms \citep{Saltelli2005}. With: 

\begin{equation}
 \sigma^2(\mathbf{y}[j](t)) = \underset{i}{\sigma}^2(\underset{-i}{E}[\mathbf{y}[j](t)|\mathbf{p}[i]]) + \underset{i}{E}[\underset{-i}{\sigma}^2(\mathbf{y}[j](t)|\mathbf{p}[i])],
 \end{equation}
the global parameter sensitivities (a.k.a. first-order Sobol' indices) are defined as:
\begin{equation}
 S^{g}[j,i](t) = \frac{\underset{i}{\sigma}^2(\underset{-i}{E}[\mathbf{y}[j](t)|\mathbf{p}[i]])}{\sigma^2(\mathbf{y}[j](t))}, \label{eq:sobolIndex}
\end{equation}
where $S^{g}(\mathbf{p}) \in \mathbf{R}^{n_y \times n_p}$, and $\sum\limits_{i=1}^{n_p}S^{g}[j,i](t)\leq1, \forall j\in\{1,\ldots,n_y\}$.

\subsection{Point Estimate Method}

To avoid a computational overload when solving the dynamic optimization problem, the Sobol' indices have to be calculated efficiently. The PEM has proven beneficial in various engineering problems \citep{Lerner02hybridbayesian}, including complex (bio)chemical and electrochemical processes \citep{Schenkendorf2018,Laue2019}. Starting with a nominal parameter vector $\mathbf{p}_0$, dedicated model parameter vector realizations $\mathbf{p}_k$ form a parameter vector set, $\mathbf{p}_k \in \mathcal{O} := \{\mathbf{p}_0,\mathcal{O}_1,-\mathcal{O}_1,\mathcal{O}_2,-\mathcal{O}_2,\mathcal{O}_3,-\mathcal{O}_3\}$, where: 
\begin{subequations}\label{eq:pem_gen}
\begin{align}
    &\nonumber\mathcal{O}_1 :=\{\mathbf{p}_0[i]+\vartheta, \; \forall i\in \{1,\ldots,n_p\}\},\\
	&\nonumber\mathcal{O}_2 :=\{\mathbf{p}_0[(i,j)]+[+\vartheta,+\vartheta], \; \forall i,\underset{j>i}{j}\in \{1,\ldots,n_p\}\},\\
	&\nonumber\mathcal{O}_3 :=\{\mathbf{p}_0[(i,j)]+[-\vartheta,+\vartheta], \; \forall i,\underset{j>i}{j}\in \{1,\ldots,n_p\}\}.
\end{align}
\end{subequations}
The overall parameter sample number, $n_{PEM}$, scales quadratically with the dimension of uncertain model parameters: 
\begin{equation}
    n_{PEM} = 2n_p^2+1 . \label{eq:pemSampleNumber}
\end{equation}
Based on the parameter samples $\mathbf{p}_k$, statistics of the output functions can be approximated. For instance, the expected value, $\mathbf{E}[\cdot]$, of the output is defined as:
 \begin{equation}\label{eq:pem_mean}
\mathbf{E}[\mathbf{y}(\mathbf{p})] \approx \sum\limits_{k=1}^{n_{PEM}}w_k \mathbf{y}(\mathbf{p}_k),
\end{equation}
where, assuming a standard Gaussian distribution, the permutation parameter and weight factors are $\vartheta = \sqrt{3}, w_0 = 1 + \frac{n_p^2 - 7n_p}{18}, w_{1,\ldots,2n_p+1} = \frac{4-n_p}{18}, w_{2n_p+2,\ldots,n_{PEM}} = \frac{1}{36}$. Note that any parametric or non-parametric probability distribution of relevant model parameters can be considered via a (non)linear transformation step, including parameter correlations \citep{xie2018robust}. 

Next, the variance, $\sigma^2[\cdot]$, can be estimated with the following equation:
\begin{equation}\label{eq:pem_var}
\sigma^2[\mathbf{y}(\mathbf{p})]  \approx \sum\limits_{k=1}^{n_{PEM}} w_k (\mathbf{y}(\mathbf{p}_k) - \mathbf{E}[\mathbf{y}(\mathbf{p})])^2 .
\end{equation}
Note that due to a nested re-sampling strategy, the global sensitivity matrix (Eq. \eqref{eq:sobolIndex}) can be determined highly efficiently with $n_{PEM}$ model simulations, where appropriate subsets, $\mathcal{P}_i \subset \mathcal{O}$, are evaluated to calculate $S^{g}(\mathbf{p}[i]),$ $\forall i \in \{1,\ldots,n_p\}$. For more details regarding the PEM in GSA and robust process design, please refer to \cite{Schenkendorf2018}.
\par

To rate the efficiency of the GSA-MBDoE, the following efficiency measure is used:
\begin{equation}\label{eq:efficiency}
    \eta=\sigma^2_l[\mathbf{\hat{p}}] / \sigma^2_g[\mathbf{\hat{p}}],
\end{equation}

where the uncertainty in the estimated model parameters, $\mathbf{\hat{p}}$, is quantified with empirical statistics: 
\begin{equation}\label{eq:mc_var}
\sigma^2_{l/g}[\mathbf{\hat{p}}]  \approx \sum\limits_{k=1}^{n_{MC}} \frac{1}{n_{MC}-1} (\mathbf{\hat{p}}_k - \mathbf{E}[\mathbf{\hat{p}}])^2.
\end{equation}
Here, $n_{MC}$ Monte Carlo simulations with artificial data assuming additive white measurement noise are used. 

\section{Case Study}

To demonstrate the effectiveness of the proposed GSA-MBDoE concept (Eq. \eqref{eq:sobolIndex}), optimal experimental operating conditions for the SPMeT (Eqs. (1)-(23)) are calculated and compared with the outcome of the standard MBDoE based on local parameter sensitivities (Eq. \eqref{eq:localSens}). In particular, we optimize an experiment with a fixed duration time, $t_{exp}$ = \SI{1000}{s}. The optimal input sequence is considered to be piece-wise constant over each \SI{100}{s} resulting in a control variable vector of $n_v=10$ elements. Each element is limited to [$\mathbf{u}[i]_{min}$,$\mathbf{u}[i]_{max}$] = [\SI{-15}{C},\SI{15}{C}], with $i=1,\,\cdots,\,n_v$. The measurement sampling time of the correspondent voltage and temperature is $t_s=\si{5}{s}$; i.e., $\mathbf{y}(t_k)=[V(t_k);T(t_k)]$. Note that considering a measurements sampling time which is independent from the number of control variables enables to adapt the DoE to the time constants of different processes.  The same initial conditions $x_0$ are used for all MBDoE results: the positive stoichiometry is initialized as $0.83$ (which corresponds to a SOC of $5\%$), and the initial temperature is set equal to \SI{298.15}{K}, while  the initial electrolyte concentration and the average concentration flux are assumed to start at equilibrium values of $1000\,$mol/m$^3$ and zero, respectively. The experiments should not exceed temperature and  voltage limits ($T(t_k)\leq\SI{320}{K}$ and $\SI{2.7}{V}\leq V(t_k)\leq\SI{4.2}{V}$), which are taken into account by the soft constraint in \eqref{equ:ineq}. For the sake of simplicity, nine normalized performance-relevant model parameters are studied, and the parameter vector reads as $\mathbf{p}=[De^0,\, E_{a,D_{p,s}},\, k_p^0,\, k_n^0,\, E_{a,k_p},\, E_{a,k_n},\, \tau_s,\, \tau_n,\,  h_{c}]$. Note that the information of all nominal model parameters can be found in \cite{Ecker2015, Ecker2015a}. When considering nine model parameters, the overall sample number needed to calculate the global parameter sensitivity matrix, $S^g$, reads as $n_{PEM}=163$. Note that for GSA, a standard deviation of 10\% is assumed for these nine model parameters. In this study, the so-called \textit{D-criteria} \citep{Walter1997} is implemented as a cost function:
\begin{equation}\label{eq:Dcriteria}
    \Phi(S^{l/g}(\mathbf{p})) = det(S^{l/g}(\mathbf{p})^TS^{l/g}(\mathbf{p})),
\end{equation}

where $F(\mathbf{p}) := S^{l}(\mathbf{p})^TS^{l}(\mathbf{p})$ can be considered as the FIM. 
\begin{figure}
\begin{center}
\includegraphics[trim= 0 0 0 0,width=8.4cm]{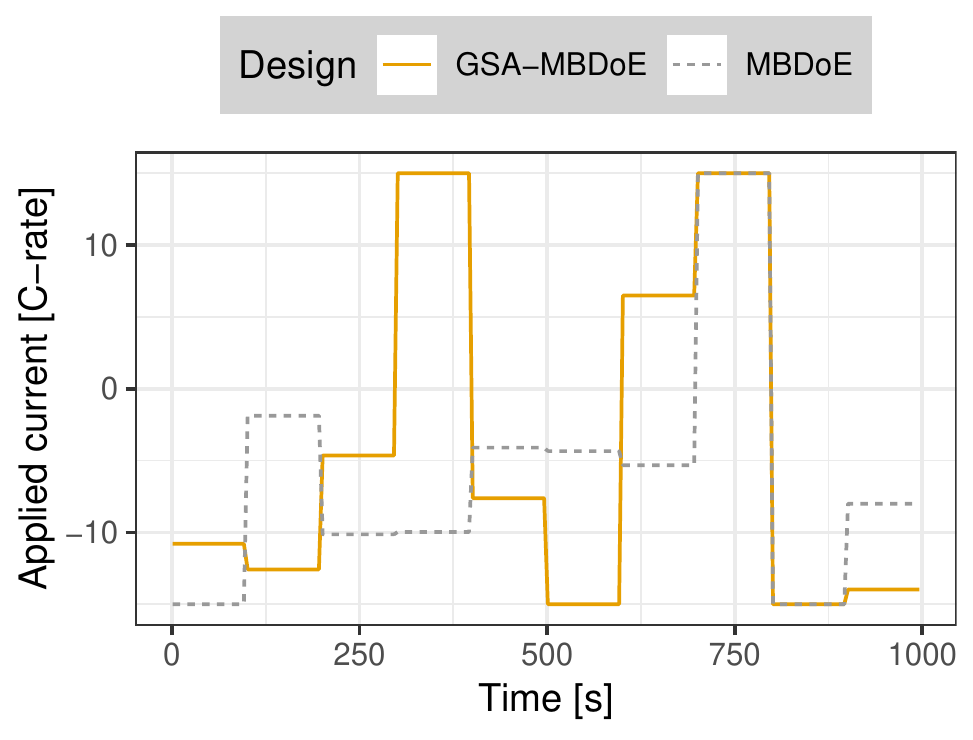}    
\caption{Current input profiles based on the MBDoE (local approach) and the GSA-MBDoE (global approach).} 
\label{fig:current}
\end{center}
\end{figure}

Technically, the resulting optimization problem (Eq. \eqref{eq:opt_str}) was solved by using the interior point NLP solver IPOPT, where a multi-start strategy was used to avoid local minima. In Fig. \ref{fig:current}, we show the optimized current input profiles obtained with the MBDoE and the GSA-MBDoE. The profiles in the first interval, $t\leq\SI{500}{s}$, show different trends. Both start with the high negative current input, but only  GSA-MBDoE switches to the high positive current input afterward. In the second interval, $t>\SI{500}{s}$, the resulting current profiles show a \textit{bang-bang} control behavior, that is, switching from the high negative current input to the high positive current input and back.

\begin{figure*}
\begin{center}
\includegraphics[trim= 0 0 0 0,width=\textwidth]{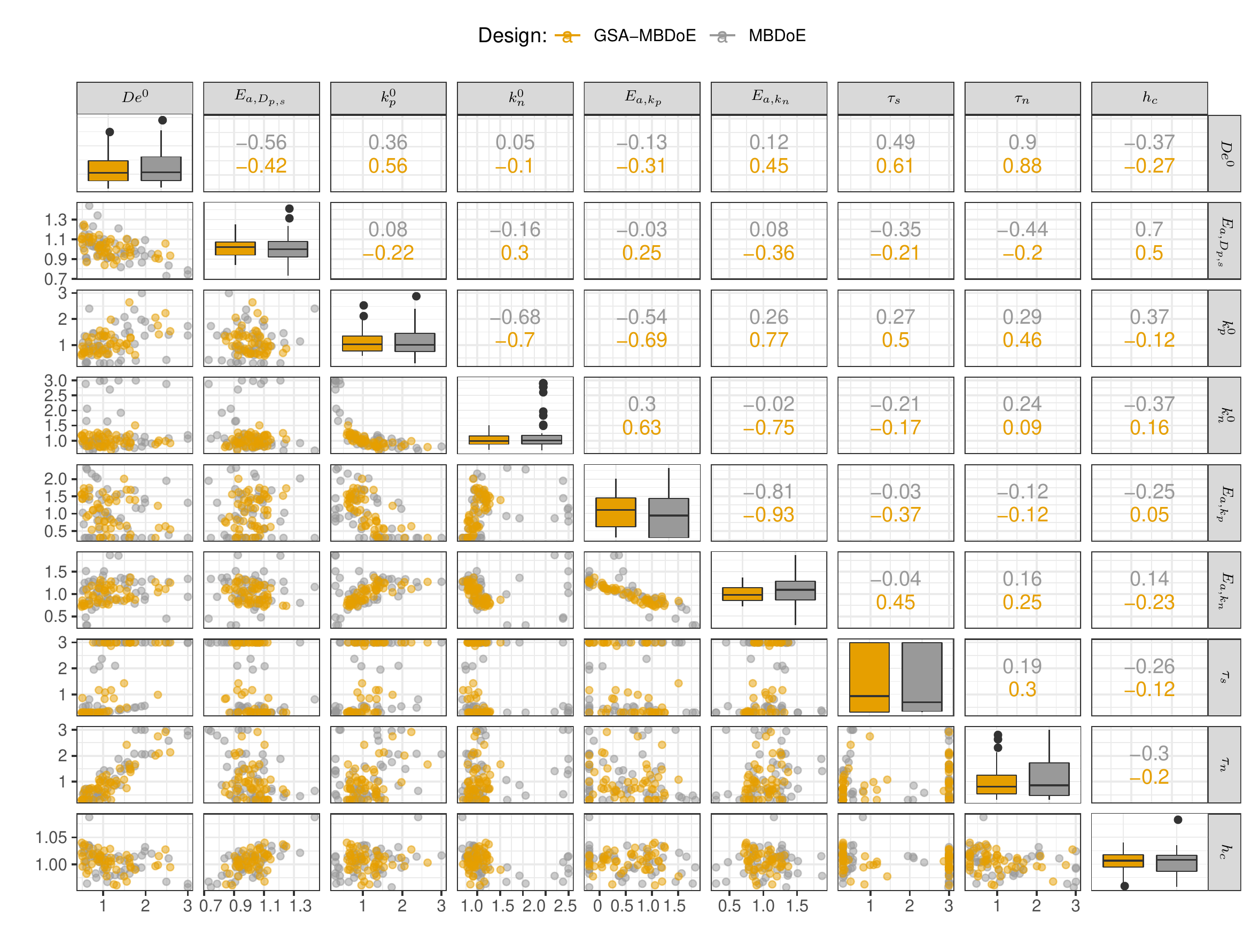}    
\caption{Statistics of the parameter estimates for the MBDoE and the GSA-MBDoE: scatter plots of all parameter combinations (lower-left triangle), corresponding box-and-whisker plots (diagonal), and parameter correlations (upper-right triangle).} 
\label{fig:scatterPlot}
\end{center}
\end{figure*}

\begin{figure*}[htb]
\centering
    \begin{subfigure}[b]{0.3\textwidth}
    \includegraphics[trim=0 0 0 0,width=1\textwidth]{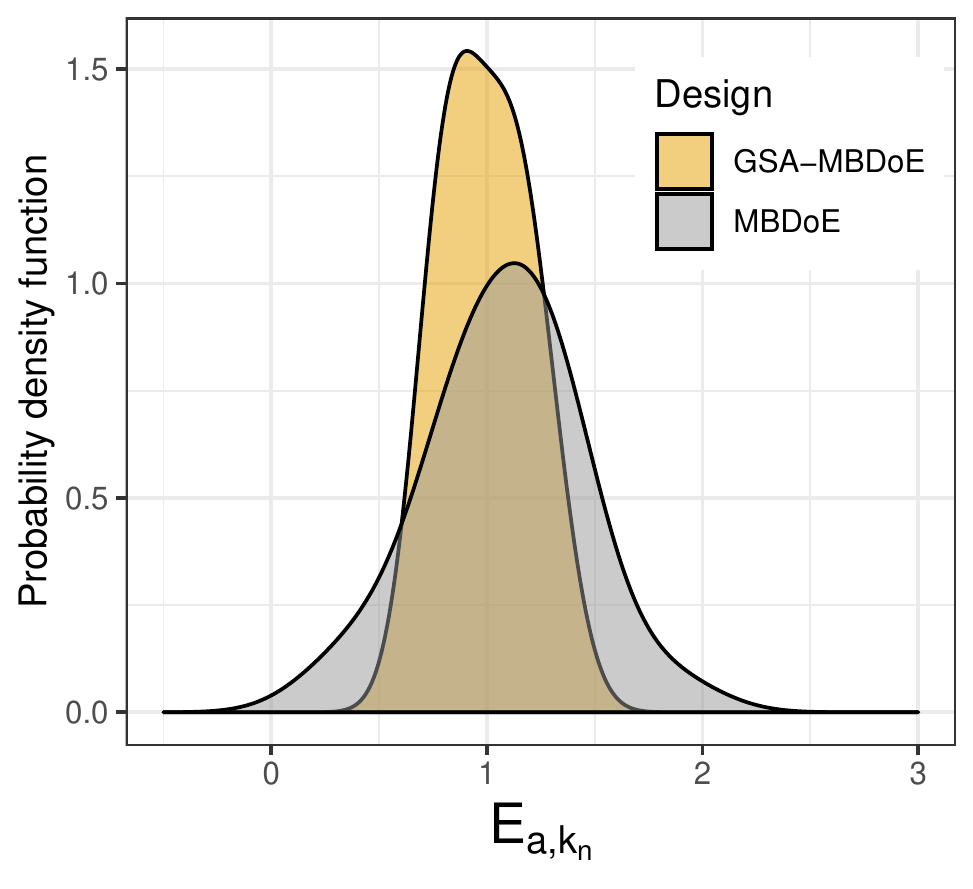}
   \caption{Gaussian}
   \label{Fig:pdf_p6}
    \end{subfigure}
    \begin{subfigure}[b]{0.3\textwidth}
    \includegraphics[trim=0 0 0 0,width=1\textwidth]{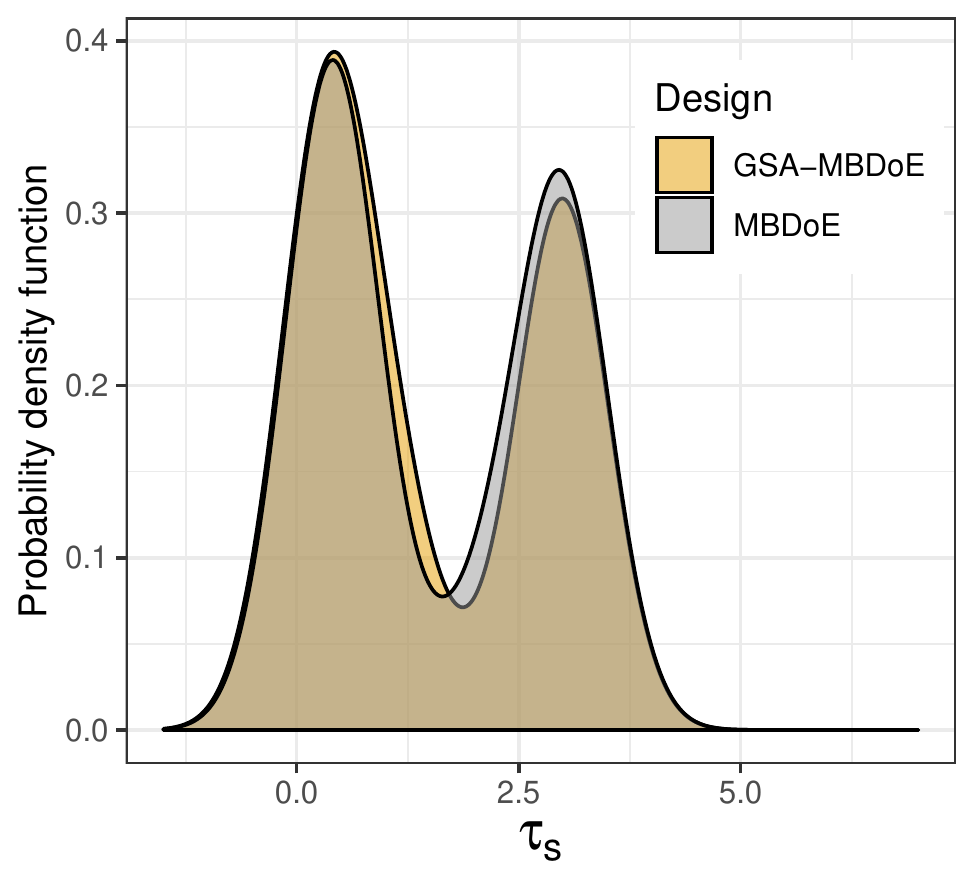}
   \caption{Bimodal}
   \label{Fig:pdf_p7}
    \end{subfigure}
    \begin{subfigure}[b]{0.3\textwidth}
    \includegraphics[trim=0 0 0 0,width=1\textwidth]{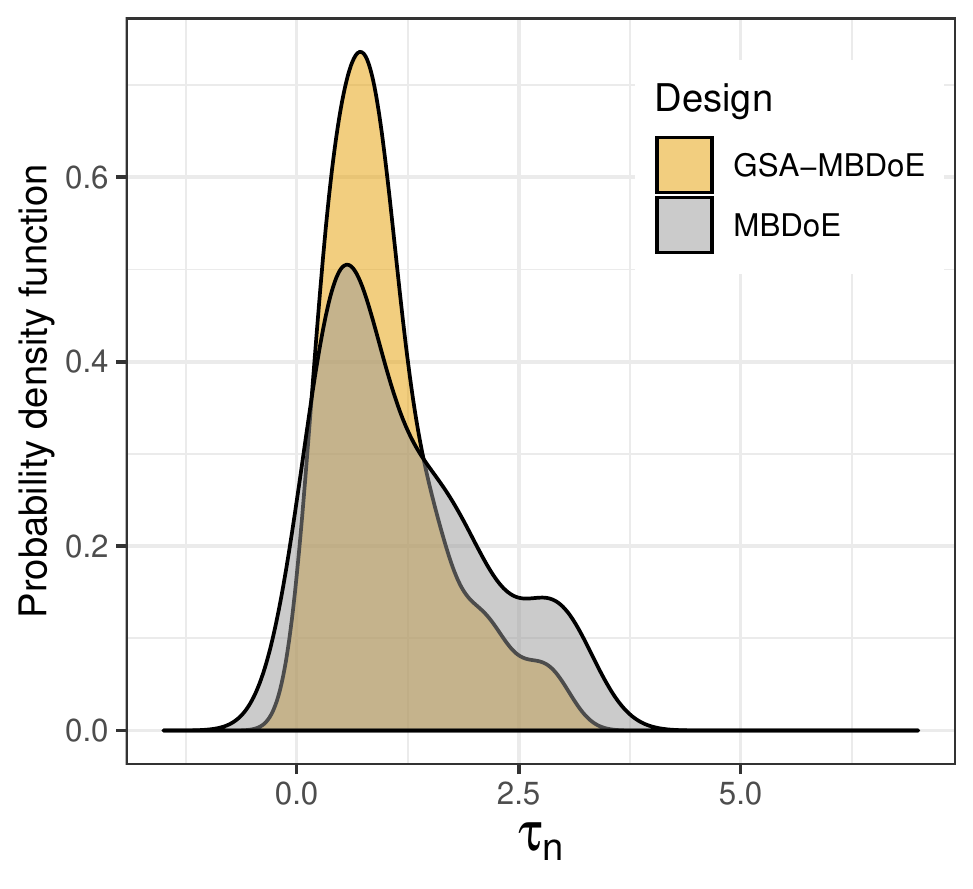}
   \caption{Non-symmetric}
   \label{Fig:pdf_p8}
    \end{subfigure}
    \caption{Resulting probability density functions of the estimated parameters based on the MBDoE and the GSA-MBDoE show: (a) Gaussian, (b) bimodal, and (c) non-symmetric distributions.}
    \label{Fig:para_pdfs}
\end{figure*}

The performance of the MBDoE and GSA-MBDoE designs is validated with Monte Carlo simulations (Eq. \eqref{eq:mc_var}), that is, 100 simulated experimental data sets with additive white noise ($\sigma^2_y(V)=10^{-2}, \sigma^2_y(T)=0.3$) are used for the parameter identification step. Based on the parameter estimates (Eq. \eqref{eq:parameterIdentification}) and the resulting parameter uncertainties (Eq. \eqref{eq:mc_var}) the efficiency of the GSA-MBDoE (Eq. \ref{eq:efficiency}), is given in Table \ref{tb:efficiency}. The efficiency measure, $\eta$, clearly shows an improvement of the GSA-MBDoE result compared to the MBDoE outcome. For all parameters, the GSA-MBDoE ensures more precise parameter estimates, that is, the optimized current input profile based on the GSA-MBDoE generates more informative data than the MBDoE did. Note that the measurement sample numbers for the GSA-MBDoE and MBDoE design are identical. The model parameter uncertainties of $k_n^0$ and $E_{a,k_n}$ are reduced significantly, and the parameter $\tau_s$ is marginally affected by a particular current input profile. In Fig. \ref{fig:scatterPlot}, we study the resulting parameter uncertainties in more detail. In the lower-left triangle, the scatter plots of all parameter combinations are given. The MBDoE results in stronger parameter variations and outliers in comparison with the GSA-MBDoE; see $k_n^0$ and $E_{a,k_n}$ results. In the case of $\tau_s$, two sample clusters can be detected, indicating two local minima of the parameter identification problem, which is insensitive to the GSA-MBDoE or MBDoE setting. On the diagonal, we illustrate the corresponding box-and-whisker plots. Obviously, the GSA-MBDoE-based parameter estimates are more precise and have fewer outliers. Based on the illustrated median and spread, only a few parameters might have a Gaussian probability density function, as in Fig. \ref{Fig:pdf_p6} for $E_{a,k_n}$.  In most cases, the probability density functions are non-Gaussian or non-symmetric, which is common for non-linear identification problems. For instance, the probability density function of $\tau_s$ is bimodal and has two peaks, because of the two local minima of the parameter identification problem; see Fig. \ref{Fig:pdf_p7}. The probability density function of $\tau_n$, in turn, shows a significant skewness in its estimates; see Fig. \ref{Fig:pdf_p8}. In the upper-right triangle of Fig. \ref{fig:scatterPlot}, the parameter correlations are shown. The GSA-MBDoE does not guarantee the lowest parameter correlation for all parameter combinations. Note, however, that parameter correlations were not included in the cost function when using the \textit{D-criteria}, but could be considered explicitly with dedicated anti-correlation criteria. 

\begin{table}[t]
\begin{center}
\caption{GSA-MBDoE efficiency according to Eq. \eqref{eq:efficiency}.}\label{tb:efficiency}
\begin{tabular}{cccccc} \hline
Parameter & $De^0$ & $E_{a,D_{p,s}}$ & $k_p^0$ & $k_n^0$ & $E_{a,k_p}$ \\\hline \vspace{0.5cm}
Efficiency $\eta$ & 1.3970 & 2.2971 & 1.9015 & 17.6513 & 1.7172 \\  \hline
Parameter & $E_{a,k_n}$ & $\tau_s$ & $\tau_n$ & $h_{c}$ & $\;$ \\\hline
Efficiency $\eta$ & 3.2228 & 1.0050 & 1.8365 & 1.6938 & $\;$ \\ \hline
\end{tabular}
\end{center}
\end{table}

\section{Conclusions}

The usefulness of model-based concepts in advanced battery management systems depends critically on the quality of the model parameters. In this work, we successfully demonstrated that a model-based experimental design, which evaluates global parameter sensitivities (GSA-MBDoE) instead of local parameter sensitivities (MBDoE), ensures informative data and more precise parameter estimates, respectively. As a case study, the  single-particle model with electrolyte and thermal dynamics (SPMeT) was implemented, and optimal current profiles were identified. Moreover, the point estimate method (PEM) ensured low computational costs of the proposed GSA-MBDoE concept. In the case of non-globally identifiable parameter identification problems, e.g., a parameter with several local minima, optimal experimental design concepts have to be advanced with rigorous parameter identifiability measures. Moreover, novel ideas of a fast global sensitivity analysis are needed if we have to study the impact of more model parameters or if we want to solve more complex optimization problems, e.g., a higher dimension of the control input vector or additional degrees of freedom of the experimental design.  


\bibliography{literature}             

\begin{thebibliography}{39}
\providecommand{\natexlab}[1]{#1}
\providecommand{\url}[1]{\texttt{#1}}
\providecommand{\urlprefix}{URL }
\expandafter\ifx\csname urlstyle\endcsname\relax
  \providecommand{\doi}[1]{doi:\discretionary{}{}{}#1}\else
  \providecommand{\doi}{doi:\discretionary{}{}{}\begingroup
  \urlstyle{rm}\Url}\fi

\bibitem[{Bizeray et~al.(2018)Bizeray, Kim, Duncan, and
  Howey}]{bizeray2018identifiability}
Bizeray, A.M., Kim, J.H., Duncan, S.R., and Howey, D.A. (2018).
\newblock Identifiability and parameter estimation of the single particle
  lithium-ion battery model.
\newblock \emph{IEEE Transactions on Control Systems Technology}, 27(5),
  1862--1877.

\bibitem[{Chaturvedi et~al.(2010)Chaturvedi, Klein, Christensen, Ahmed, and
  Kojic}]{Chaturvedi2010}
Chaturvedi, N.A., Klein, R., Christensen, J., Ahmed, J., and Kojic, A. (2010).
\newblock Algorithms for advanced battery-management systems.
\newblock \emph{IEEE Control Systems}, 30(3), 49--68.

\bibitem[{Chu and Hahn(2013)}]{Chu2013}
Chu, Y. and Hahn, J. (2013).
\newblock {Necessary condition for applying experimental design criteria to
  global sensitivity analysis results}.
\newblock \emph{Computers \& Chemical Engineering}, 48, 280--292.

\bibitem[{Di~Domenico et~al.(2010)Di~Domenico, Stefanopoulou, and
  Fiengo}]{DiDomenico2010}
Di~Domenico, D., Stefanopoulou, A., and Fiengo, G. (2010).
\newblock Lithium-ion battery state of charge and critical surface charge
  estimation using an electrochemical model-based extended {Kalman} filter.
\newblock \emph{Journal of Dynamic Systems, Measurement, and Control}, 132(6),
  061302.

\bibitem[{Doyle et~al.(1993)Doyle, Fuller, and Newman}]{Doyle1993}
Doyle, M., Fuller, T.F., and Newman, J. (1993).
\newblock Modeling of galvanostatic charge and discharge of the
  lithium/polymer/insertion cell.
\newblock \emph{Journal of the Electrochemical Society}, 140(6), 1526--1533.

\bibitem[{Ecker et~al.(2015{\natexlab{a}})Ecker, K{\"a}bitz, Laresgoiti, and
  Sauer}]{Ecker2015a}
Ecker, M., K{\"a}bitz, S., Laresgoiti, I., and Sauer, D.U.
  (2015{\natexlab{a}}).
\newblock Parameterization of a physico-chemical model of a lithium-ion battery
  {II. Model} validation.
\newblock \emph{Journal of The Electrochemical Society}, 162(9), A1849--A1857.

\bibitem[{Ecker et~al.(2015{\natexlab{b}})Ecker, Tran, Dechent, K{\"a}bitz,
  Warnecke, and Sauer}]{Ecker2015}
Ecker, M., Tran, T.K.D., Dechent, P., K{\"a}bitz, S., Warnecke, A., and Sauer,
  D.U. (2015{\natexlab{b}}).
\newblock Parameterization of a physico-chemical model of a lithium-ion battery
  {I. Determination} of parameters.
\newblock \emph{Journal of The Electrochemical Society}, 162(9), A1836--A1848.

\bibitem[{Forman et~al.(2012)Forman, Moura, Stein, and
  Fathy}]{forman2012genetic}
Forman, J.C., Moura, S.J., Stein, J.L., and Fathy, H.K. (2012).
\newblock Genetic identification and fisher identifiability analysis of the
  doyle--fuller--newman model from experimental cycling of a lifepo4 cell.
\newblock \emph{Journal of Power Sources}, 210, 263--275.

\bibitem[{Gomadam et~al.(2002)Gomadam, Weidner, Dougal, and
  White}]{Gomadam2002}
Gomadam, P.M., Weidner, J.W., Dougal, R.A., and White, R.E. (2002).
\newblock Mathematical modeling of lithium-ion and nickel battery systems.
\newblock 110(2), 267--284.

\bibitem[{Hu et~al.(2012)Hu, Li, and Peng}]{hu2012comparative}
Hu, X., Li, S., and Peng, H. (2012).
\newblock A comparative study of equivalent circuit models for li-ion
  batteries.
\newblock \emph{Journal of Power Sources}, 198, 359--367.

\bibitem[{Kiefer(1959)}]{Kiefer1959}
Kiefer, J. (1959).
\newblock Optimum experimental designs.
\newblock \emph{Journal of the Royal Statistical Society. Series B
  (Methodological)}, 21(2), 272--319.

\bibitem[{Laue et~al.(2019)Laue, Schmidt, Dreger, Xie, R{\"{o}}der,
  Schenkendorf, Kwade, and Krewer}]{Laue2019}
Laue, V., Schmidt, O., Dreger, H., Xie, X., R{\"{o}}der, F., Schenkendorf, R.,
  Kwade, A., and Krewer, U. (2019).
\newblock {Model-Based Uncertainty Quantification for the Product Properties of
  Lithium-Ion Batteries}.
\newblock \emph{Energy Technology}, 1900201.

\bibitem[{Lerner(2002)}]{Lerner02hybridbayesian}
Lerner, U.N. (2002).
\newblock Hybrid bayesian networks for reasoning about complex systems.
\newblock Technical report.

\bibitem[{Manesso et~al.(2017)Manesso, Sridharan, and Gunawan}]{Gunawan2017}
Manesso, E., Sridharan, S., and Gunawan, R. (2017).
\newblock Multi-objective optimization of experiments using curvature and
  fisher information matrix.
\newblock \emph{Processes}, 5(4).

\bibitem[{Mendoza et~al.(2016)Mendoza, Rothenberger, Hake, and
  Fathy}]{mendoza2016optimization}
Mendoza, S., Rothenberger, M., Hake, A., and Fathy, H. (2016).
\newblock Optimization and experimental validation of a thermal cycle that
  maximizes entropy coefficient fisher identifiability for lithium iron
  phosphate cells.
\newblock \emph{J. Power Sources}, 308, 18--28.

\bibitem[{Mendoza et~al.(2017)Mendoza, Rothenberger, Liu, and
  Fathy}]{mendoza2017maximizing}
Mendoza, S., Rothenberger, M., Liu, J., and Fathy, H.K. (2017).
\newblock Maximizing parameter identifiability of a combined thermal and
  electrochemical battery model via periodic current input optimization.
\newblock \emph{IFAC PapersOnLine}, 50(1), 7314--7320.

\bibitem[{Moura(2015)}]{Moura2015}
Moura, S.J. (2015).
\newblock Estimation and control of battery electrochemistry models: A
  tutorial.
\newblock In \emph{2015 54th IEEE Conference on Decision and Control (CDC)},
  3906--3912. IEEE.

\bibitem[{Moura et~al.(2017)Moura, Argomedo, Klein, Mirtabatabaei, and
  Krstic}]{Moura2017}
Moura, S.J., Argomedo, F.B., Klein, R., Mirtabatabaei, A., and Krstic, M.
  (2017).
\newblock Battery state estimation for a single particle model with electrolyte
  dynamics.
\newblock \emph{IEEE Transactions on Control Systems Technology}, 25(2),
  453--468.

\bibitem[{Ning and Popov(2004)}]{Ning2004}
Ning, G. and Popov, B.N. (2004).
\newblock Cycle life modeling of lithium-ion batteries.
\newblock \emph{Journal of The Electrochemical Society}, 151(10), A1584--A1591.

\bibitem[{Park et~al.(2018{\natexlab{a}})Park, Kato, Gima, Klein, and
  Moura}]{park2018b}
Park, S., Kato, D., Gima, Z., Klein, R., and Moura, S. (2018{\natexlab{a}}).
\newblock Optimal experimental design for parameterization of an
  electrochemical lithium-ion battery model.
\newblock \emph{Journal of The Electrochemical Society}, 165(7), A1309--A1323.

\bibitem[{Park et~al.(2018{\natexlab{b}})Park, Kato, Gima, Klein, and
  Moura}]{park2018a}
Park, S., Kato, D., Gima, Z., Klein, R., and Moura, S. (2018{\natexlab{b}}).
\newblock Optimal input design for parameter identification in an
  electrochemical li-ion battery model.
\newblock In \emph{2018 Annual American Control Conference (ACC)}, 2300--2305.
  IEEE.

\bibitem[{Perez et~al.(2016)Perez, Hu, and Moura}]{Perez2016}
Perez, H.E., Hu, X., and Moura, S.J. (2016).
\newblock Optimal charging of batteries via a single particle model with
  electrolyte and thermal dynamics.
\newblock 4000--4005.

\bibitem[{Perez et~al.(2017)Perez, Hu, Dey, and Moura}]{Perez2017}
Perez, H.E., Hu, X., Dey, S., and Moura, S.J. (2017).
\newblock Optimal charging of {Li}-ion batteries with coupled
  electro-thermal-aging dynamics.
\newblock \emph{IEEE Transactions on Vehicular Technology}, 66(9), 7761--7770.

\bibitem[{Pozzi et~al.(2018{\natexlab{a}})Pozzi, Ciaramella, Gopalakrishnan,
  Volkwein, and Raimondo}]{Pozzi2018c}
Pozzi, A., Ciaramella, G., Gopalakrishnan, K., Volkwein, S., and Raimondo, D.M.
  (2018{\natexlab{a}}).
\newblock Optimal design of experiment for parameter estimation of a single
  particle model for lithiumion batteries.
\newblock In \emph{2018 IEEE Conference on Decision and Control (CDC)},
  6482--6487. IEEE.

\bibitem[{Pozzi et~al.(2018{\natexlab{b}})Pozzi, Ciaramella, Volkwein, and
  Raimondo}]{Pozzi2018}
Pozzi, A., Ciaramella, G., Volkwein, S., and Raimondo, D.M.
  (2018{\natexlab{b}}).
\newblock Optimal design of experiments for a lithium-ion cell: {P}arameters
  identification of an isothermal single particle model with electrolyte
  dynamics.
\newblock \emph{Industrial \& Engineering Chemistry Research}, 58(3),
  1286--1299.

\bibitem[{Pukelsheim(2006)}]{pukelsheim2006optimal}
Pukelsheim, F. (2006).
\newblock \emph{Optimal design of experiments}.
\newblock SIAM.

\bibitem[{Rao and Swift(2006)}]{Rao2006}
Rao, M.M. and Swift, R.J. (2006).
\newblock \emph{Probability theory with applications}.
\newblock Springer.

\bibitem[{Rodriguez-Fernandez et~al.(2007)Rodriguez-Fernandez, Kucherenko,
  Pantelides, and Shah}]{Fernandez2007}
Rodriguez-Fernandez, M., Kucherenko, S., Pantelides, C., and Shah, N. (2007).
\newblock Optimal experimental design based on global sensitivity analysis.
\newblock \emph{Computer Aided Chemical Engineering}, 24, 63 -- 68.

\bibitem[{Saltelli et~al.(2005)Saltelli, Ratto, Tarantola, and
  Campolongo}]{Saltelli2005}
Saltelli, A., Ratto, M., Tarantola, S., and Campolongo, F. (2005).
\newblock {Sensitivity analysis for chemical Models}.
\newblock \emph{Chemical Reviews}, 105, 2811--2828.

\bibitem[{Santhanagopalan et~al.(2006)Santhanagopalan, Guo, Ramadass, and
  White}]{Santhanagopalan2006a}
Santhanagopalan, S., Guo, Q., Ramadass, P., and White, R.E. (2006).
\newblock Review of models for predicting the cycling performance of lithium
  ion batteries.
\newblock \emph{Journal of Power Sources}, 156(2), 620--628.

\bibitem[{Schenkendorf et~al.(2018)Schenkendorf, Xie, Rehbein, Scholl, and
  Krewer}]{Schenkendorf2018}
Schenkendorf, R., Xie, X., Rehbein, M., Scholl, S., and Krewer, U. (2018).
\newblock {The Impact of Global Sensitivities and Design Measures in
  Model-Based Optimal Experimental Design}.
\newblock \emph{Processes}, 6(4), 27.

\bibitem[{{Scire Jr.} et~al.(2001){Scire Jr.}, Dryer, and
  Yetter}]{ScireJr.2001}
{Scire Jr.}, J., Dryer, F., and Yetter, R. (2001).
\newblock {Comparison of global and local sensitivity techniques for rate
  constants determined using complex reaction mechanisms}.
\newblock \emph{International Journal of Chemical Kinetics}, 33(12), 784--802.

\bibitem[{Sinkoe and Hahn(2017)}]{Hahn2017}
Sinkoe, A. and Hahn, J. (2017).
\newblock Optimal experimental design for parameter estimation of an il-6
  signaling model.
\newblock \emph{Processes}, 5(3).

\bibitem[{Subramanian et~al.(2005)Subramanian, Diwakar, and
  Tapriyal}]{Subramanian2005}
Subramanian, V.R., Diwakar, V.D., and Tapriyal, D. (2005).
\newblock Efficient macro-micro scale coupled modeling of batteries.
\newblock \emph{Journal of The Electrochemical Society}, 152(10), A2002--A2008.

\bibitem[{Torchio et~al.(2016)Torchio, Magni, Gopaluni, Braatz, and
  Raimondo}]{Torchio2016}
Torchio, M., Magni, L., Gopaluni, R.B., Braatz, R.D., and Raimondo, D.M.
  (2016).
\newblock Lionsimba: {A} matlab framework based on a finite volume model
  suitable for {Li}-ion battery design, simulation, and control.
\newblock \emph{Journal of The Electrochemical Society}, 163(7), A1192--A1205.

\bibitem[{Turanyi(1990)}]{Turanyi1990}
Turanyi, T. (1990).
\newblock {Sensitivity Analysis of Complex Kinetic Systems. Tools and
  Applications}.
\newblock \emph{Journal of Mathematical Chemistry}, 5, 203--248.

\bibitem[{Walter and Pronzato(1997)}]{Walter1997}
Walter, E.E. and Pronzato, L. (1997).
\newblock \emph{{Identification of parametric models from experimental data}}.
\newblock Springer.

\bibitem[{Xie et~al.(2018)Xie, Krewer, and Schenkendorf}]{xie2018robust}
Xie, X., Krewer, U., and Schenkendorf, R. (2018).
\newblock Robust optimization of dynamical systems with correlated random
  variables using the point estimate method.
\newblock \emph{IFAC-PapersOnLine}, 51(2), 427--432.

\bibitem[{Zou et~al.(2014)Zou, Manzie, and Anwar}]{Zou2014}
Zou, C., Manzie, C., and Anwar, S. (2014).
\newblock Control-oriented modeling of a lithium-ion battery for fast charging.
\newblock \emph{IFAC Proceedings Volumes}, 47(3), 3912--3917.

\end{thebibliography}

\end{document}